\documentclass[%
 prl,
superscriptaddress,
 amsmath,amssymb,
 aps,twocolumn
]{revtex4-2}
\usepackage{epsfig} 
\usepackage{bmpsize}
\usepackage{mathtools}
\usepackage{color}

\newcommand{\bra}[1]{\left< #1 \right|}
\newcommand{\ket}[1]{\left| #1 \right>}
\newcommand{\mv}[1]{\left< #1 \right>}
\newcommand{\+}{\dagger}

\begin{document} 
 
 \title{Spin-Triplet Pairing in Heavy Nuclei is Stable Against Deformation}

\author{Georgios Palkanoglou}
\affiliation{Department of Physics, University of Guelph, 
Guelph, Ontatio N1G 2W1, Canada}
\affiliation{TRIUMF, 4004 Wesbrook Mall, Vancouver, British Columbia V6T 2A3, Canada}
\author{Michael Stuck}
\affiliation{Department of Physics, University of Guelph, 
Guelph, Ontatio N1G 2W1, Canada}
\author{Alexandros Gezerlis}
\affiliation{Department of Physics, University of Guelph, 
Guelph, Ontatio N1G 2W1, Canada}

\begin{abstract} 
Any experimental evidence of nucleons paired in spin-triplet states will confirm the existence of an exotic phase of nuclear systems. This type of nuclear superfluidity has been hypothesized in heavy nuclei, where the antagonizing spin-orbit effects are damped, and there it oftentimes coexists with traditional spin-singlet pairing, leading to the possibility of mixed-spin pairing. Realistic nuclear deformation, not considered in such studies, could make-or-break these proposals, since its effect on triplet pairing, and the competition (and coexistence) of the two superfluid phases, was expected to be crucial. We report on a thorough study on the effect of deformation on triplet, singlet, and mixed-spin pairing in the relevant region of the nuclear chart. We find that, at low isospin asymmetries, spin-triplet pairing is enhanced by deformation, while below the proton-drip line, the novel superfluid phase survives alongside the usual spin-singlet pairing. These results suggest that spin-triplet superfluidity exists in realistic nuclei and can be probed in the lab.
\end{abstract} 

\maketitle 
{\textit{Introduction}} --- Spin-triplet neutron-proton pairs have been elusive, with the only pairing seen in experimentally accessible nuclei being of spin-singlet and between like particles~\cite{Frauendorf:2014}. In contrast, our understanding of nuclear forces tells of an increased attraction in the spin-triplet channel of the two-body nuclear interaction, underlined by the existence of the deuteron. This has generated a long-standing question mark on the existence of deuteronlike spin-triplet pairs in nuclei. At a mean-field level, the strong spin-orbit field has been identified as the suppressor of the spin-triplet superfluidity in nuclei~\cite{Poves:1998,Bertsch:2011}, which suggests the possible existence of the exotic superfluid phase in heavy nuclei and extended nuclear systems. At large, nuclear superfluidity remains a vibrant field probed with many approaches ~\cite{Soma:2011, Hinohara:2016, Tichai:2018, Schuck:2019,Drissi:2021, Barbieri:2021,Palkanoglou:2022,Obiol:2023,Hinohara:2023}, with ties to astrophysics~\cite{Martin:2016,Sourie:2021,Rios:2017,Sedrakian:2019} and nuclear reactions~\cite{Niu:2018, Magierski:2017, Bernard:2019, Oishi:2017}.

Robust spin-triplet pairing has been suggested to exist close to the $N=Z$ line of the nuclear chart, $A\sim 130$, with a region of an exotic mixed-spin condensate lying in the crossover between spin-triplet and spin-singlet pairing~\cite{Bertsch:2011,Gezerlis:2011,Frauendorf:2014}. This region of the nuclear chart is characterized by a finite deformation~\citep{Moller:2016}, so far neglected in Refs.~\cite{Bertsch:2011,Gezerlis:2011, Bulthuis:2016, Rrapaj:2019}, which could change this picture dramatically: with the two pairing phases in general working against each other, and deformation typically antagonizing pairing altogether~\cite{Gambacurta:2015}, the total effect is nontrivial to predict.
Furthermore, the existence of spin-triplet pairing in the nuclear ground state relies on the proximity of low-$j$ shells to the Fermi surface that happens in the $A=130$ region; that too will be altered by deformation. In this Letter, we address nuclear deformation, an important missing piece which can greatly modify pairing correlations: we confirm the existence of spin-triplet pairs in heavy nuclei, at $N=Z$, but also within the physical region, where it coexists with spin-singlet pairing, and can be probed in the lab.
\vfill\null

{\textit{Context}} --- We use a generalization of the Hamiltonian used in Refs.~\cite{Bertsch:2011,Gezerlis:2011, Bulthuis:2016, Rrapaj:2019}, which in second quantization reads
\begin{equation}
H=\sum_{ij}\epsilon_{ij}c_i^\+ c_{j} + \sum_{i>j, k>l} \bra{ij}v\ket{kl} c_i^\+  c_j^\+ c_l c_k~. \label{eq:ham}
\end{equation}
The one-body part, $\epsilon$, contains the kinetic energy, a deformed Wood-Saxon well, and a deformed spin-orbit term. At the mean-field level the external Wood-Saxon potential models the shape of the nucleus’ surface and the spin-orbit term, peaking at the nuclear surface, is defined on the same shape:
\begin{align}
\epsilon = \frac{p^2}{2m} + V_{\rm WS}(\varpi,z;\vec{\alpha})+ 
C\boldsymbol{\nabla}\left[ V_{\rm WS}(\varpi, z;\vec{\alpha})\right]\cdot \left(\boldsymbol{\sigma} \times \mathbf{p}\right)~. \label{eq:sp}
\end{align}
The last term in Eq.~(\ref{eq:sp}) is the generalized spin-orbit form that reduces to the familiar $(1/r)dV_{\rm WS}/dr\,\left(\mathbf{l}\cdot\mathbf{s}\right)$ in the absence of deformation~\cite{Bender:2003,Teran:2002,Teran:2003}; it is multiplied by its coupling, $C$~\cite{BohrMottelson:book}.
In lieu of experimental results in the mass region of $A\sim 130$, we follow the predictions of Ref.~\cite{Moller:2016} that prescribe axial deformation for that region with a moderate quadrupole component and small higher-order ones.

The specific shape of the nucleus’s surface is given by a vector of deformation parameters, namely, $\vec{\alpha}=(\varepsilon, \vec{\beta})$. This contains the elongation of an underlying Cassini oval, $\varepsilon$, and the additional deformation described on a basis of spherical harmonics with coefficients $\beta_\lambda$ truncated to some degree $\lambda_{\rm max}$~\cite{Stravinsky:1968,Pashkevich:1971}. The vector $\vec{\alpha}$ uniquely defines the nuclear surface,
\begin{align}
    \phi(\varpi,z) &= \left[\left(\bar{z}^2+\bar{\varpi}^2\right)^2-2\varepsilon R_0^2\left(\bar{z}^2-\bar{\varpi}^2\right)+\varepsilon^2R_0^4\right]^{1/4} - \notag \\
    & -R_0 \left[1+\sum_{\lambda=1}^{\lambda_{\rm max}}\beta_\lambda Y_{\lambda 0}(x)\right]=0~,
\end{align}
and the one-body potential this generates,
\begin{align}
    V_{\rm WS}(\varpi,z) &= V_0 \left[1+\exp \left(\frac{1}{a}\frac{\phi}{|\boldsymbol \nabla\phi|}\right)\right]^{-1}~.
\end{align}
The barred coordinates, $\bar{\varpi}=f(R,x;\varepsilon)$ and $\bar{z}=g(R,x;\varepsilon)$ are cylindrical coordinates parametrized by the curvilinear coordinates $(R,x)$ in which Cassini ovals span coordinate surfaces and $R_0$ and $a$ are the radius and surface diffuseness, respectively, of the nondeformed nucleus.

The two-body part of the Hamiltonian, $v$, is a contact pairing interaction,
\begin{equation}
\bra{ij}v\ket{kl} = \sum_\alpha v_\alpha \bra{ij} \delta^{(3)}\left(\mathbf{x}-\mathbf{x}'\right)  P_\alpha P_{J_z=0} \ket{kl} ~,\label{eq:int}
\end{equation}
where the projection operator $P_\alpha$ projects on six spin-isospin channels: $\alpha=0,1,2$ correspond to the three spin-singlet and isospin-triplet channels, while $\alpha=3,4,5$ correspond to the three spin-triplet and isospin-singlet channels. The projection operator $P_{J_z=0}$ takes advantage of the axial symmetry and pairs particles in time reversed states, resulting in axially symmetric pair-wavefunctions. It conveniently generalizes the common approximation of pairs with zero orbital angular momentum\cite{Bertsch:2011,Frauendorf:2014} allowing formation of pairs with finite angular momentum which are expected to form in ground states of deformed nuclei. Similar pairing schemes have been used before in the study of spin-singlet and spin-triplet nuclear pairing~\cite{Sandulescu:2015,Negrea:2018}. 
The interaction strengths that appear in Eq.~(\ref{eq:int}) are tuned to reproduce earlier works~\cite{Bertsch:2011,Gezerlis:2011,Bulthuis:2016,Rrapaj:2019} which corresponds to the ratio $v_t/v_s=1.38$. Nonetheless, our conclusions are not sensitive to the value of the ratio (see the Supplemental Material~\cite{suppl}).

We allow for six pairing channels, the three spin-singlet and three spin-triplet ones defined in Eq.~(\ref{eq:int}). Channels with the same total spin (or isospin) and different projections are prescribed the same $v_\alpha$ and are seen as equivalent, that is, we use an interaction with only two interaction strengths.

{\textit{HFB treatment and its application}} --- The Hartree-Fock-Bogolyubov (HFB) treatment of the Hamiltonian in Eq.~(\ref{eq:ham}) corresponds to identifying the Bogolyubov transformation that minimizes the ground-state energy. The system is then described as a collection of free quasiparticles, whose vacuum defines the ground state. In practice, the Bogolyubov transformation is parametrized by the matrices $U$ and $V$, which mix particle and hole operators in Fock space defining the quasiparticles. These matrices also define the normal and anomalous densities, $\rho=V^*V^T$ and $\kappa = V^\star U^T$, respectively, which are fundamental to the description, as they in turn define the HFB expectation value of the Hamiltonian,
\begin{equation}
H^{00}=\textrm{Tr}\left[\epsilon \rho  - \frac{1}{2} \Delta \kappa^*\right]~, \label{eq:h00}
\end{equation}
where $\Delta_{ij} = \frac{1}{2}\sum_{kl}v_{ijkl} \kappa_{kl}$.

The HFB treatment can be implemented very effectively via the gradient method outlined in Ref.~\cite{Robledo:2011}, and used in a similar fashion as in Refs.~\cite{Bulthuis:2016,Gezerlis:2011,Bertsch:2011}: we locate the HFB minimum by performing a gradient descent on the surface of the HFB energy parametrized by the $U$ and $V$ matrices (or $\kappa$) in Eq. (\ref{eq:h00}). In this way one can straightforwardly identify solutions of different symmetries by constraining the descent, i.e., introducing the appropriate terms with Lagrange multipliers to Eq. (\ref{eq:h00}).  We use 8 constraining fields: the neutron and proton numbers and six pairing amplitudes that are defined below. This allows us to explore different HFB minima, apart from the global minimum that corresponds to the fully paired ground state.

The spin character of an HFB state can be described by the expectation value of the pair operator $\bar{P}_\alpha=P_\alpha P_{J_z=0}$ that defines the six pairing channels and appears in Eq.~(\ref{eq:int}). These six pairing amplitudes can then be used as constraining fields to locate states with certain pairing character. Specifically, we use the state where all six pairing channels are closed, i.e., all pairing amplitudes vanish, as a reference state: we measure all energies relative to its unpaired energy, namely, $E_0$, and call the resulting quantity correlation energy, $E_{\rm corr}=E-E_0$.

{\textit{Results}} --- We mainly characterize our results by the strength and spin character of their pairing. We quantify the strength of the pairing correlations via the correlation energy. We label the spin character of the condensate using the expectation values of the pairing fields $K_\alpha=\mv{\bar{P}_\alpha}$. We define~\cite{Gezerlis:2011}
\begin{align}
    A_{\rm singlet} = \frac{\sum_{\beta=0,1,2}K^2_{\beta}}{\sum_{\alpha}K^2_{\alpha}}~, \label{eq:a0}
\end{align}
and a nucleus is labeled as spin singlet if $A_{\rm singlet} > 0.8$ and spin triplet if $A_{\rm singlet} < 0.2$, otherwise, the nucleus is labeled as mixed spin. Finally, the spin-orbit field is expected to play a major role in the interplay of the different pairing phases, and we quantify it via its expectation value at the HFB ground state:
\begin{align}
    B_{\textrm{HFB}} = \bra{\textrm{HFB}}C\nabla f(\varpi, z;\vec{\alpha})\cdot \left(\boldsymbol{\sigma} \times \mathbf{p}\right)\ket{\textrm{HFB}}~. \label{eq:so}
\end{align}

\begin{figure*}[ht]
\begin{center}
\includegraphics[width=0.9\linewidth]{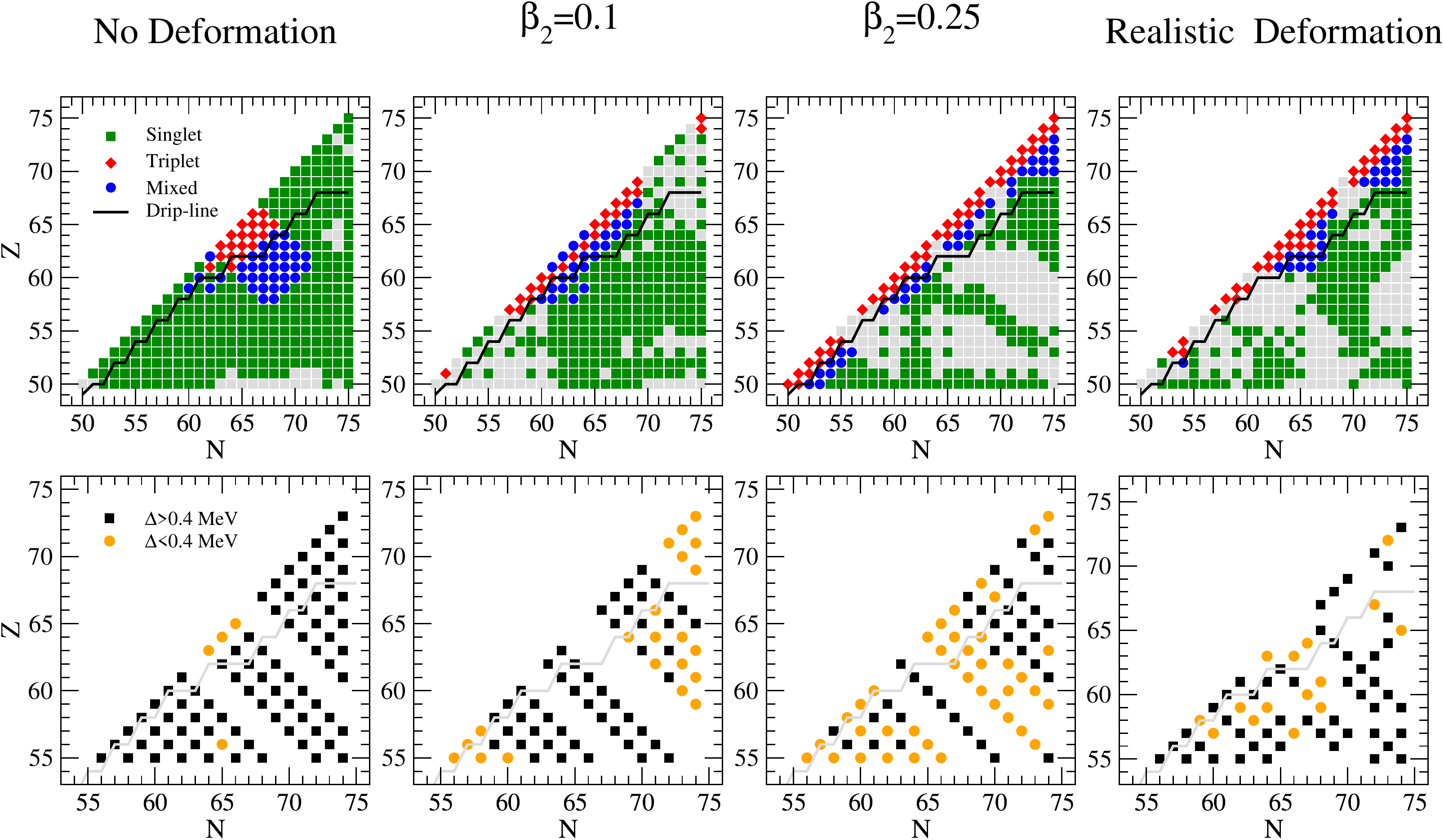}
\caption{Chart of nuclides with $Z \le N$ for neutron numbers from 50 to 75. In the top panels, we show the spin character of the nuclear condensate: the green squares, blue circles, and red diamonds, represent nuclei with singlet, mixed-spin, and triplet pairing, respectively, and the gray squares denote nuclei with $E_{\rm corr}<0.5~\textrm{MeV}$.
In the bottom panels, we show strong (black squares) and weak (orange circles) pairing gaps (see Eq. (\ref{eq:gap})) for the same deformation. The proton-drip line is drawn in black in the top panels and in gray in the bottom panels~\cite{Moller:2016}.}
\label{fig:chart}
\end{center}
\end{figure*}

We calculate correlation energies and pairing amplitudes for all nuclei in the interesting region of mass $A\sim 130$ and we plot the results in the top panels of Fig.~\ref{fig:chart}. To visualize how the condensates respond to nuclear deformation, we perform these calculations for a sequence of deformation parameters, starting from spherical shapes across the region in the leftmost panel, which reproduces the results of Refs.~\cite{Bertsch:2011,Gezerlis:2011}, increasing quadrupole deformation for all nuclei progressively in the two middle panels, and reaching the realistic deformation prescribed by M{\"o}ller \textit{et al}~\cite{Moller:2016} in the rightmost panel. The bottom panels of Fig.~\ref{fig:chart} show pairing gaps, i.e., binding energies of pairs, for the region of interest and for the same deformation parameters. These will be discussed later.

\begin{figure}[t]
\begin{center}
\begin{tabular}{c}
\includegraphics[width=0.9\columnwidth]{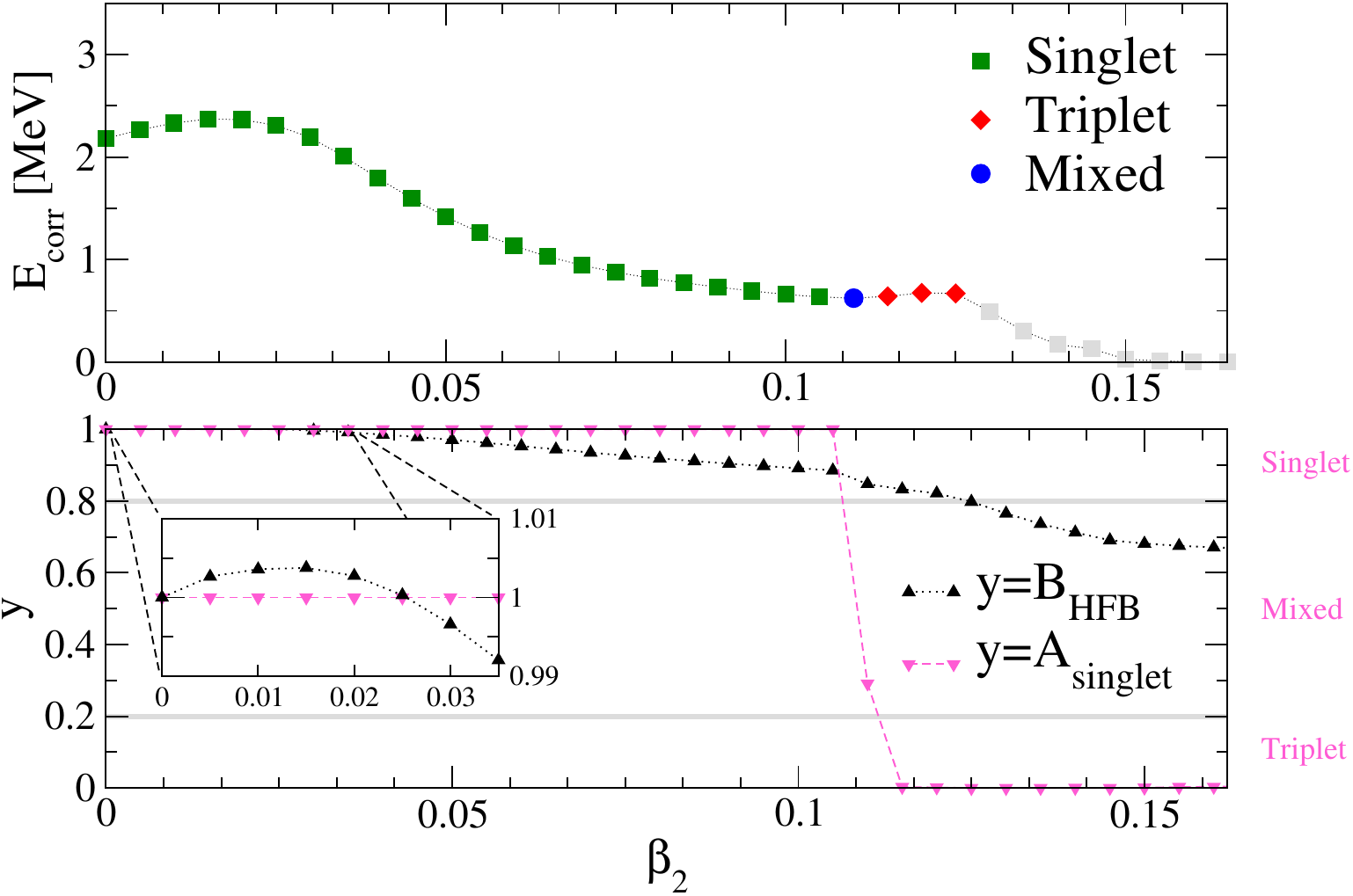}
\end{tabular}
\caption{The evolution of $\prescript{108}{54}{\textrm{Xe}}$'s ground state increasing quadrupole deformation. The top panel shows the correlation energy of the nucleus. In the lower panel, $A_\textrm{singlet}$ measures the size of the singlet condensate, relative to the triplet condensate [see Eq.~(\ref{eq:a0})] and $B_{\rm HFB}$ quantifies the spin-orbit field [see Eq.(\ref{eq:so})]. }
\label{fig:xe}
\end{center}
\end{figure}

\begin{figure}[t]
\vspace{0.5cm}
\begin{center}
\includegraphics[width=0.9\linewidth]{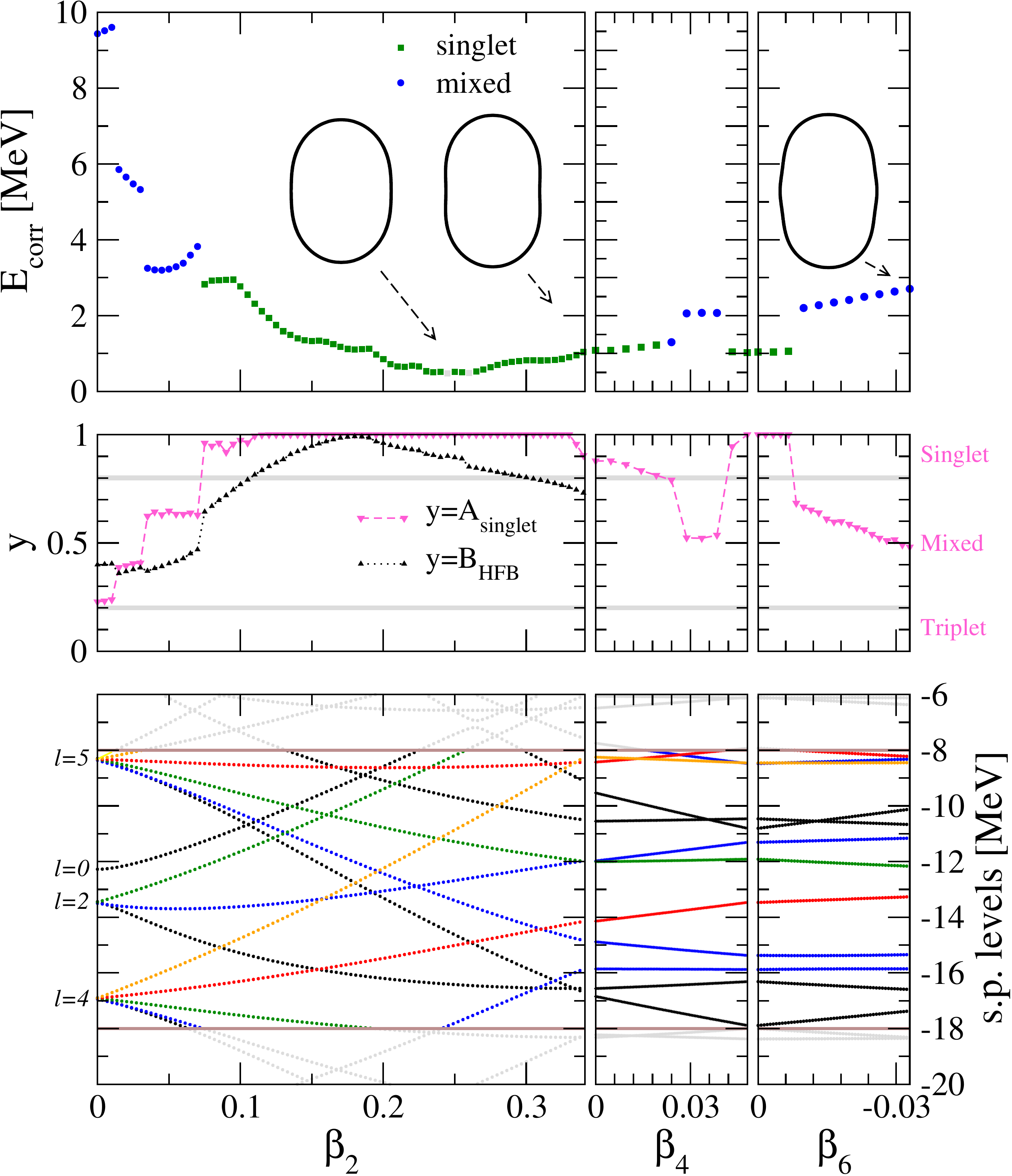}
\caption{The evolution of $\prescript{126}{61}{\textrm{Pm}}$'s ground state from no deformation to the realistic deformation predicted by Ref.~\cite{Moller:2016}. The bottom panel shows the evolution of the single-particle levels that lie in the energy window: from bottom to top $l_z=0$ (black), 1 (blue), 2 (green), 3 (red), 4 (orange), and 5 (yellow). The middle panel shows the singlet pairing amplitude overlaid with the expectation value of the spin-orbit field (at the ground-state), appropriately normalized. The top panel shows the correlation energy of the nucleus along with the shape of the nuclear surface along the evolution.}
\label{fig:pm}
\end{center}
\end{figure}

The superfluidity's response to deformation is rich so we start from the $N=Z$ line, where neutrons and protons occupy the same orbitals, increasing the odds for neutron-proton pairing. First, we investigate quadrupole deformation ($\beta_2$), which seems to drive the qualitative effects for most nuclei. At very small $\beta_2$ deformation, spin-triplet pairing is generally damped, resulting in a small increase in the correlation energies of singlet nuclei and the change of some triplet-paired nuclei to mixed spin. This can be seen at $A\sim125$ in Fig.~\ref{fig:chart} as well as the low-deformation end of Fig.~\ref{fig:xe}. As quadrupole deformation approaches the realistic value predicted by Ref.~\cite{Moller:2016} for the region, spin triplet completely dominates the $N=Z$ line inducing a mixed-spin band at $N-Z=2$ or $3$ before the nuclear condensate turns spin singlet. This dramatic change resembles the results of Ref.~\cite{Bulthuis:2016}, where the spin-orbit interaction was switched-off artificially: the enhancement of spin-triplet pairing can be understood as driven by a general suppression of the spin-orbit field at the ground state. This is clearly demonstrated in Fig.~\ref{fig:pm} for $\prescript{126}{61}{\textrm{Pm}}$ where the spin-singlet component and the spin-orbit field show similar behavior. The connection can also be seen in Fig.~\ref{fig:xe} where we plot the correlation energy, spin-singlet pairing amplitude, and spin-orbit field for $\prescript{108}{54}{\textrm{Xe}}$, a case of a pure spin-singlet paired ground state at the spherical limit turned to spin triplet by deformation. There the transition between the two pairing phases is accompanied by a steady decrease of the spin-orbit field.

On the $N=Z$ line, a reduced spin-orbit field corresponds to a spin-triplet paired ground state~\cite{Bertsch:2011, Bulthuis:2016}. Moving along an isobar (i.e., increasing $N-Z$) the nuclear condensate very quickly changes to spin singlet. This is expected, as increasing isospin asymmetry translates to increasing mismatch in the proton and neutron Fermi surfaces, making the pairing of a neutron and a proton increasingly unlikely.

Switching on the higher deformation modes, to the values predicted by Ref.~\cite{Moller:2016}, we find that for most nuclei, qualitatively, the picture does not change: spin-triplet pairing dominates at and around $N=Z$, mixed-spin nuclei can be found at small isospin asymmetry $N-Z$, and spin singlet generally takes over for higher isospin asymmetries. Additionally, the crossover region, between the two phases has been drastically reduced compared to the spherical limit. At a triangular region, close to $N=Z$, at $A\sim 130$ some triplet and mixed-spin paired nuclei reemerge, driven by higher-order deformation, with some isotopes of Pm ($Z=61$) also being inside the physical region. Specifically, the nuclei $\prescript{125}{61}{\textrm{Pm}}$, $\prescript{126}{61}{\textrm{Pm}}$, and $\prescript{127}{61}{\textrm{Pm}}$ pose an interesting case as they lie within the proton-drip line and maintain a sizable spin-triplet condensate at full deformation. These nuclei also exhibit a comparable spin-singlet condensate, coexisting with the spin triplet, making them of mixed-spin character.

In the survey of the region presented in Fig.~\ref{fig:chart} we identify some Pm isotopes that lie in the physical region, i.e., within the proton-drip line and can support spin-triplet pairing in their ground states. With all three nuclei parametrized with similar realistic deformation parameters, we plot the evolution of $\prescript{126}{61}{\textrm{Pm}}$ in Fig.~\ref{fig:pm} where it is seen that it is the higher-order deformation that drives this transition. Plotted in the same figure is also the evolution of the single-particle levels (bottom panel) demonstrating how the overall initial reduction in the correlation energy is driven by the reduction of the single-particle degeneracy close to the Fermi surface; the latter lies close to the middle of the indicated energy window. Finally, overlayed on the top panel, are the shapes of the nuclear surfaces at various deformations demonstrating a simple connection between correlation energy (or pairing) and the geometry of the nucleus.

The source of the spin-orbit field’s decrease needs a
deeper discussion. This suppression is especially counterintuitive
since the spin-orbit field is a surface effect and
deformation is moving the nuclear surface away from the
spherical shape, which has the lowest surface-to-volume
ratio, for a given volume. Indeed, perturbative calculations,
for very small quadrupole deformation, show that
an increase in the spin-orbit strength is expected, if the
nuclear surface is adequately probed by the participating
orbitals. A typical example of such an increase is seen in
the low-$\beta_2$ region of Figs.~\ref{fig:xe} and \ref{fig:pm} where in the latter we also observe the $l = 4,5$
orbitals approaching the Fermi surface
(i.e., the chemical potential). Hence, the effect of the spin-orbit field depends on whether deformation
brings higher-$j$ orbitals close to the nucleus’
Fermi surface. This is because high-$j$ wave functions lie
closer to the nuclear surface and can probe surface effects. It is important to clarify that these effects are
not a result of pairing, as they are seen both in the HFB and unpaired states.

{\textit{Outlook}} --- The correlation energies are a valuable tool in visualizing pairing correlations since they resemble the binding energy that a nucleus gains from the pairing correlations. However, they are \textit{not} experimentally observable. Therefore, we also calculated pairing gaps, i.e.,  odd-even staggering, traditionally the smoking gun of superfluidity in nuclear systems~\cite{Frauendorf:2014,Palkanoglou:2020},
\begin{align}
    \Delta(N) = E(N) - \frac{1}{2}\left[E(N-1)+ E(N+1)\right]~, \label{eq:gap}
\end{align}
where $N$ is either a proton or a neutron, and an odd number, while the other nuclear species is kept to a fixed even number~\cite{Duguet:2001}.
We plot the results in the bottom panel of Fig.~\ref{fig:chart}. We distinguish between gaps larger and smaller than $0.4$ MeV, while we have excluded results with systematic errors generated by the existence of an energy window (i.e., a hard cutoff) in the single-particle spectrum (see Fig.~\ref{fig:pm}). That way, the rightmost panel of the lower row in Fig.~\ref{fig:chart} adds nuance to the conventional wisdom dictating that deformation generally suppresses pairing, showing deformation of higher multipoles strengthening pairing correlations~\cite{Palkanoglou:prep}. Finally, Refs.~\cite{Bertsch:2011,Gezerlis:2011}, demonstrated that spin-triplet paired ground states yielded reduced pairing gaps, but Fig.~\ref{fig:chart} shows that deformation partially lifts this suppression, as was accurately foreseen in Ref.~\cite{Frauendorf:2014}.

This Letter constitutes the first systematic study of the interplay between spin-singlet pairing, spin-triplet pairing, and deformation in nuclei and it demonstrates that the deformation's effect on superfluidity can be understood as ternary. First, it weakens all types of pairing correlations by reducing the single-particle density of states, as has been shown in the past. Second, it increases the surface-to-volume ratio of a given nucleus amplifying the spin-orbit field's effect, which in turn tends to suppress spin-triplet pairing. Finally, by changing the single-particle level structure close to the Fermi surface, it might induce pairing of particles in low-$j$ orbitals, whose wave functions do not lie very close to the nucleus's surface, thus lifting the spin-orbit field's suppression and enhancing spin-triplet pairing. The interplay of the three effects, coupled with the well-documented antagonism of the two superfluid phases, gives rise to an unexpected outcome in the mass region of $A\sim130$: realistic deformation creates a spin-triplet superfluid in the ground states of most nuclei with $N=Z$, and on some nuclei with small isospin asymmetry where it almost always coexists with a spin-singlet superfluid phase. The spin-triplet phase, be it pure or mixed, was found to be moderately strong, yielding correlation energies of $\sim 2~\textrm{MeV}$ for realistically deformed nuclei. Our results predict the existence of neutron-proton spin-triplet pairs in heavy nuclei and invite further accurate yet costly \textit{ab initio} calculations~\cite{Tichai:2018, Hergert:2018,Soma:2021,Hu:2024}: the HFB ground states represented in Fig.~\ref{fig:chart} capture both deformation and neutron-proton correlations and as such they can be an appropriate reference state for an \textit{ab initio} study providing the relevant static correlations. Additionally, the existence of exotic pairing phases in heavy deformed nuclei may lead to other interesting phenomena, e.g., in nuclear dynamics, fission and fusion~\cite{Scamps:2012,Bernard:2019,Magierski:2017,Sadhukhan:2014}, or nuclear structure~\cite{Baran:2020,Sambataro:2021,Sambataro:2023} and reactions~\cite{Niu:2018} and can be significant for experiments as we have identified some nuclei on or below the proton-drip line, which can be accessed by experiment most easily, with a sizeable mixed-spin condensate when deformed~\cite{Sagawa:2016,Matsubara:2015,Isacker:2021,Frauendorf:2001}.

We thank  M.~Drissi, P.~Garrett, L.~Jokiniemi, A.~Kwiatkowski, E.~Leistenschneider, P.~Navratil, and T.~Papenbrock for useful discussions. This work was supported by the Natural Sciences and Engineering Research Council
(NSERC) of Canada and the Canada Foundation for
Innovation (CFI). TRIUMF receives federal funding via a contribution agreement with the National Research Council of Canada. Computational
resources were provided by SHARCNET and NERSC.

\end{document}


\title{Supplemental Material for: Spin-Triplet Pairing in Heavy Nuclei is Stable Against Deformation}

\author{Georgios Palkanoglou}
\affiliation{Department of Physics, University of Guelph, 
Guelph, ON N1G 2W1, Canada}
\affiliation{TRIUMF, 4004 Wesbrook Mall, Vancouver, BC V6T 2A3, Canada}
\author{Michael Stuck}
\affiliation{Department of Physics, University of Guelph, 
Guelph, ON N1G 2W1, Canada}
\author{Alexandros Gezerlis}
\affiliation{Department of Physics, University of Guelph, 
Guelph, ON N1G 2W1, Canada}

\maketitle 

\begin{figure}[H]
\centering
\begin{subfigure}[b]{0.32\textwidth}
   \includegraphics[width=1\linewidth]{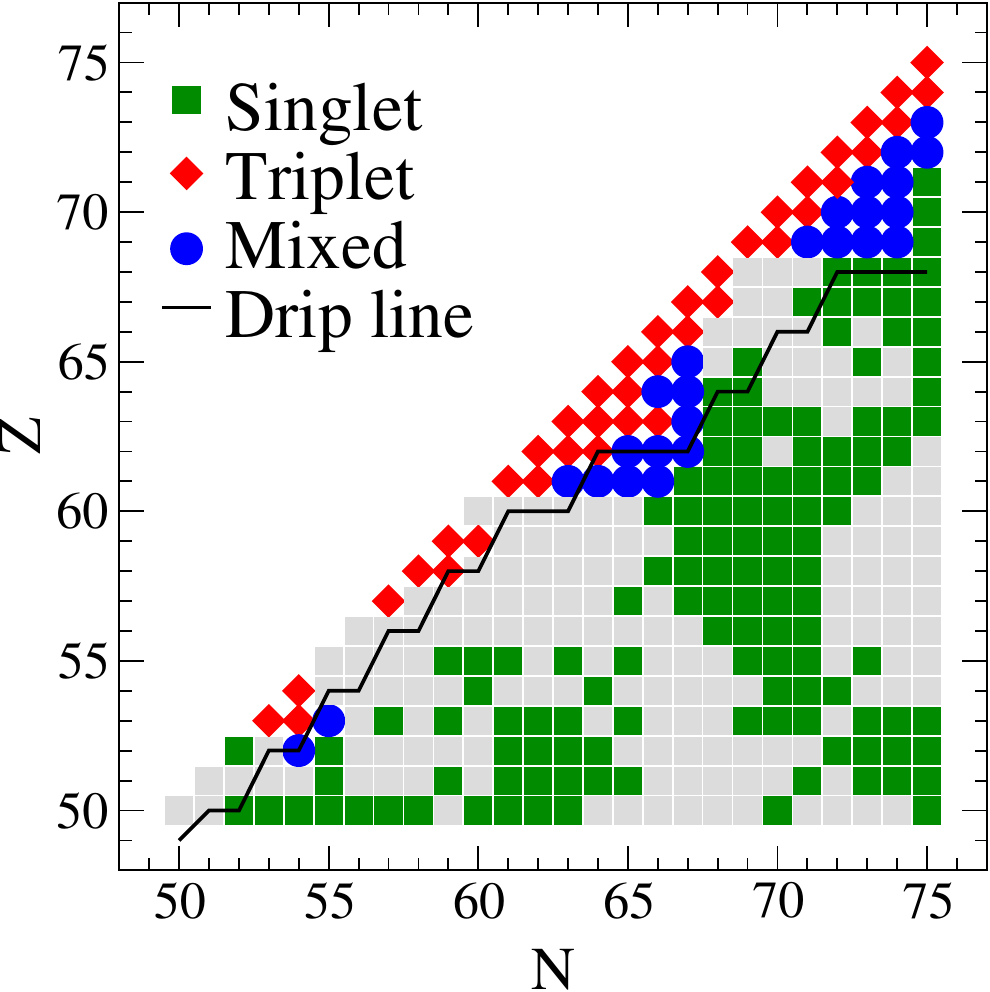}
\end{subfigure}
\begin{subfigure}[b]{0.32\textwidth}
   \includegraphics[width=1\linewidth]{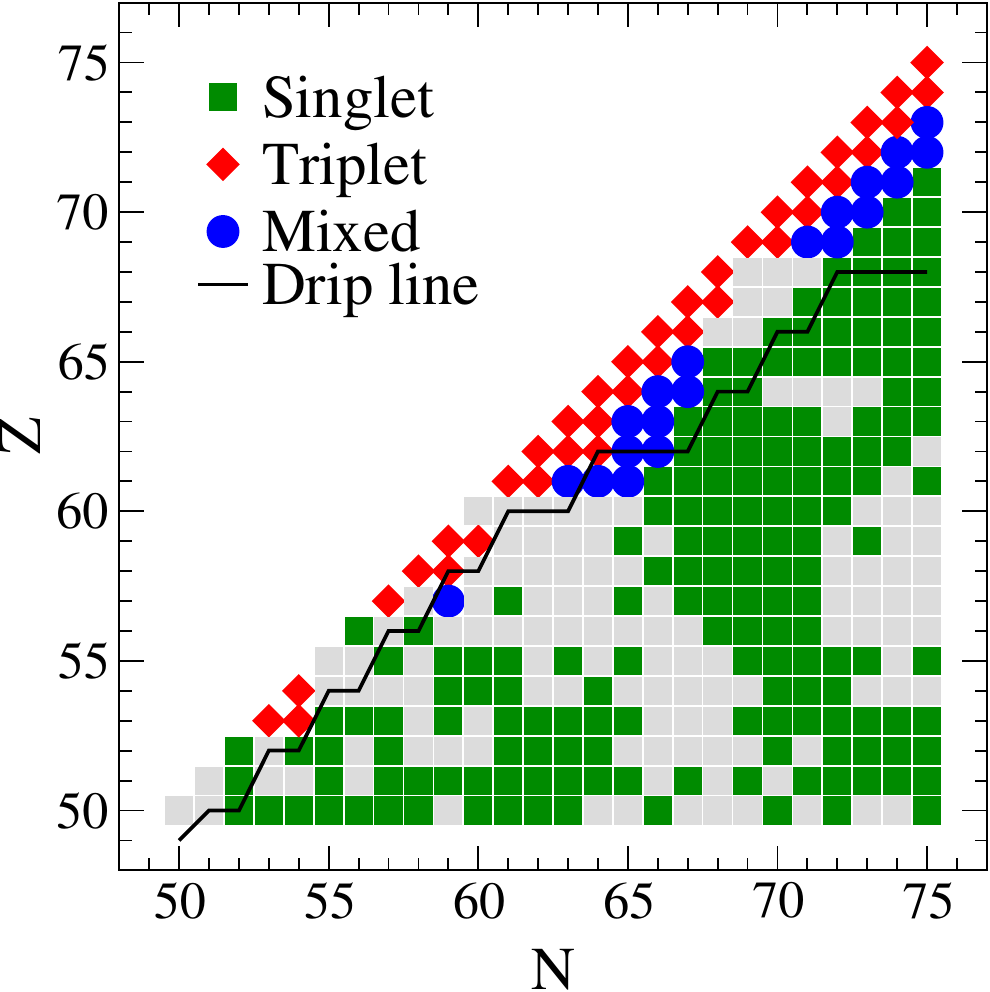}
\end{subfigure}
\caption{ The condensates in the mass region $A\sim 130$ for $v_t/v_s=1.3$ (bottom) and $v_t/v_s=1.41$ (top).\label{fig:chart}}
\end{figure}

The Hamiltonian used in this Letter, 
\begin{align}
    H=\sum_{ij}\epsilon_{ij}c_i^\+ c_{j} + \sum_{i>j, k>l} \bra{ij}v\ket{kl} c_i^\+  c_j^\+ c_l c_k~, \label{eq:ham}
\end{align}
includes a contact pairing interaction of the form
\begin{align}
    \bra{ij}v\ket{kl} = \sum_\alpha v_\alpha \bra{ij} \delta^{(3)}\left(\mathbf{x}-\mathbf{x}'\right)  P_\alpha P_{J_z=0} \ket{kl}~.
\end{align}
This is active in six spin-isospin channels: three spin-singlet and isospin-triplet channels ($\alpha=0,1,2$) with interaction strength $v_s$ and three spin-triplet and isospin-singlet channels ($\alpha=3,4,5$) with interaction strength $v_t$. That is,
\begin{align}
    v_\alpha=v_s~,&\quad \alpha=0,1,2,\\
    v_\alpha=v_t~,&\quad \alpha=3,4,5~.
\end{align}
This means that the pairing interaction's strength doesn't differentiate between different $T=1$ pairs in the singlet channel and between different $S=1$ pairs in the triplet channel. The values $v_s$ and $v_t$ are chosen to reproduce the correlation energies and pairing amplitudes of the HFB calculations in the spherical limit in Refs.~\cite{Bertsch:2011,Gezerlis:2011,Bulthuis:2016,Rrapaj:2019}. This fit yields  $v_t/v_s=1.38$ and to verify that our main results are insensitive to that value we performed a series of calculations varying it. In Fig.~\ref{fig:chart} we present the condensate's spin-character in the ground states of nuclides in the $A\sim130$ mass region at realistic deformation for $v_t/v_s=1.3$ (bottom) and $v_t/v_s=1.41$ (top). 
We compare this to the top right-most panel of Fig.~1 in the main Letter and confirm that our qualitative conclusions are stable: spin-triplet pairing
overtakes the $N = Z$ line with two ``islands'' of spin-triplet paired nuclei protruding from the $N=Z$
line centered at $Z \approx 62$ and $Z \approx 72$, and mixed-spin pairing still extends in the physical region. 

Finally, we have compared the pairing gaps of Fig. 1 in the main Letter with pairing gaps extracted from the AME collaboration's mass evaluation~\cite{ame:2020}. In the more neutron-rich part of the $A\sim 130$ region, and where the 
 \noindent AME data is available, we have verified that we reproduce the general trend of increasing gaps with isospin asymmetry.

